\documentclass[a4paper,12pt]{article}
\usepackage[a4paper,left = 2cm, right= 2cm,top = 2cm, bottom = 2cm]{geometry}
\usepackage{cite}
\usepackage{amsmath,amssymb,amsfonts}
\usepackage{multirow}
\usepackage{algorithmic}
\usepackage{graphicx}
\usepackage{textcomp}
\usepackage{amssymb,
amsthm,
amsfonts,
amsmath,
enumerate,
caption,
subfigure,diagbox,booktabs,bm,array,tikz,cuted
}
\allowdisplaybreaks
\usepackage[ruled,vlined,commentsnumbered]{algorithm2e}
\newcommand{\me}{\mathrm{e}}

\newtheorem{defi}{Definition} 
\newtheorem{thm}{Theorem}
\newtheorem{corollary}{Corollary}
\newtheorem{lem}{Lemma}
\newtheorem{pf}{Proof}

\usepackage{authblk}

\begin{document}
\bibliographystyle{ieeetr}
\title{Investigating the Convergence of Sigmoid-Based Fuzzy General Grey Cognitive Maps}
\date{}
\author[a]{Xudong Gao} 
\author[a]{Xiaoguang Gao} 
\author[b]{Jia Rong} 
\author[c]{Xiaolei Li}
\author[d]{Ni Li}
\author[e]{Yifeng Niu}
\author[a,f]{Jun Chen\thanks{Corresponding Author, e-mail:junchen@nwpu.edu.cn}}

\affil[a]{School of Electronics and Information, Northwestern Polytechnical University, Xi'an 710072, Shaanxi, China}
\affil[b]{Department of Data Science and AI, Monash University, Clayton, Melbourne,VIC3800, Victoria, Australia}
\affil[c]{School of Electrical Engineering,
Yanshan University, Qinhuangdao 066000, Hebei, China}
\affil[d]{School of Aeronautics, Northwestern Polytechnical University, Xi'an 710072, Shaanxi, China}
\affil[e]{College of Intelligence Science and Technology, National University of Defense Technology, Changsha, 410073, Hunan, China}
\affil[f]{Chongqing Institute for Brain and Intelligence, Guangyang Bay Laboratory, Nan'an District, 400064, Chongqing, China}

\newcommand\blfootnote[1]{
\begingroup
\renewcommand\thefootnote{}\footnote{#1}%
\addtocounter{footnote}{-1}
\endgroup
}

\maketitle

\begin{abstract}
  The Fuzzy General Grey Cognitive Map (FGGCM) and Fuzzy Grey Cognitive Map (FGCM) extend the Fuzzy Cognitive Map (FCM) by integrating uncertainty from multiple interval data or fuzzy numbers. Despite extensive studies on the convergence of FCM and FGCM, the convergence behavior of FGGCM under sigmoid activation functions remains underexplored. This paper addresses this gap by deriving sufficient conditions for the convergence of FGGCM to a unique fixed point. 
  Using the Banach and Browder-Gohde-Kirk fixed point theorems, and Cauchy-Schwarz inequality, the study establishes conditions for the kernels and greyness of FGGCM to converge to unique fixed points. A Web Experience FCM is adapted to design an FGGCM with weights modified to GGN. Comparisons with existing FCM and FGCM convergence theorems confirm that they are special cases of the theorems proposed here.  
  The conclusions support the application of FGGCM in domains such as control, prediction, and decision support systems.

\end{abstract}

\section{Introduction}  \label{sec:introduction}
The fuzzy cognitive map (FCM), introduced by Kosko in 1986, is composed of interconnected nodes and weighted links between them \cite{KOSKO1986}. Its architecture closely resembles that of a recurrent neural network, employing activation functions such as sigmoid or tanh to facilitate reasoning. Unlike traditional neural networks, FCM offers superior interpretability \cite{Napoles2023}. This is attributed to its straightforward structure, with nodes having explicit meanings and no hidden layers, allowing it to simulate the cognitive processes of the human brain. Over the past four decades, FCM has been effectively utilized across a wide range of fields \cite{Lei2024,Wang2024,Zhe2020,Luo2023,Li2024}.

During the iterative reasoning process of FCMs, the values of their nodes can converge to fixed points, limit cycles, or chaotic states. Understanding the convergence behavior of FCMs is crucial for applications such as large-scale system simulation, the development of learning algorithms, and pattern recognition. Ambiguous convergence can create significant obstacles in FCM design, potentially causing the system to settle into undesired or chaotic states, thereby undermining its effectiveness. Consequently, the convergence properties of FCMs have increasingly become a focal point of scholarly investigation.

The study of FCM convergence began with efforts to use Lyapunov functions for stability analysis, though these methods proved inapplicable due to FCM's complex feedback connections \cite{Kosko1996}. Subsequent work identified FCM's three convergence scenarios—fixed points, limit cycles, and chaotic states—and explored how parameters influence transitions between these behaviors. Limited data was shown to often lead to limit cycle states \cite{Tsadiras1999, Taber2001, Taber2007}. Conditions for FCM convergence to unique or non-unique fixed points were derived using the Banach fixed point theorem, with applications extending to power plant control, fault detection, and time series prediction \cite{Boutalis2009, Karatzinis2018, Karatzinis2018a, Karatzinis2018b, Karatzinis2021, Karatzinis2021a}. Later research demonstrated that the number of fixed points in sigmoidal FCMs depends on \(\lambda\), a parameter that controls the steepness of sigmoid function, where smaller values ensure a single stable point, while larger values allow multiple solutions \cite{Napoles2014}. Clear parameter boundaries were introduced to guarantee unique fixed points, and organizational models were analyzed to identify attractors and chaotic behaviors \cite{Knight2014, Hatwagner2017}. Recent advances include conditions for global stabilization, parameter ranges to reduce output uncertainty, and convergence criteria for multi-valued FCMs \cite{Harmati2023a, Koutsellis2022, Maximov2023}. FCMs were also shown to converge to limit state spaces rather than fixed points under certain conditions, with methods proposed for quantifying biases and analyzing attractor uniqueness \cite{Concepcion2021, Napoles2022}. Semi-tensor product techniques have further enabled global stability conditions, attractor determination, and controller design \cite{Xiaojie2022}. These studies clarified the convergence states of FCMs and their influencing factors, providing a solid theoretical foundation for understanding FCM dynamics. 
Sufficient conditions for convergence to fixed points under different activation functions and parameters enable the design of stable models with proven applications across various domains. However, analyzing the impact of multi-interval data uncertainty on FCM convergence requires further exploration.

The convergence characteristics of FCMs are crucial for their design and application. Fixed-point convergence reflects system stabilization, applied in ecosystem modeling, energy-efficient controls, and intelligent teaching systems \cite{Peng2016, Behrooz2019, Song2024}. Limit cycle convergence models periodic phenomena like climate change awareness \cite{Biloslavo2012}. However, avoiding chaotic states is essential in practical. Convergence insights also drive learning algorithm development. Methods like Active Hebbian Learning and cultural algorithms optimize weight matrices and enhance stability \cite{Papageorgiou2005, Ahmadi2014}. Hybrid supervised-unsupervised approaches improve topology and global convergence, while Particle Swarm Optimization methods and node-specific activation functions boost efficiency and speed \cite{Altundogan2018, Napoles2016, Napoles2017, Napoles2018}. These advancements underline the importance of convergence in refining FCMs for diverse applications.

Extensions to FCMs have been proposed to improve uncertainty representation and dynamic reasoning. Fuzzy General Grey Cognitive Map (FGGCM), an extension of FCM and Fuzzy Grey cognitive maps (FGCM), incorporates general grey numbers (GGN) from grey system theory, enabling the handling of multiple interval data through iterative reasoning based on GGN's kernel and degree of greyness \cite{Chen2020}. Similar to FCM and FGCM, FGGCM typically employs tanh and sigmoid activation functions in continuous scenarios. When node values fall within the range \([-1, 1]\), tanh is used as the activation function, whereas sigmoid is applied when the range is \([0, 1]\). Due to the differing computation methods for tanh and sigmoid functions in the context of GGN, particularly in the calculation of greyness, the convergence conditions of FGGCM also vary. This study focuses specifically on the convergence conditions of FGGCM when the activation function is sigmoid. Recent studies have explored FGCM convergence conditions. Analyses of FCMs and FGCMs with sigmoid and tanh activation functions have identified sufficient conditions for fixed-point convergence \cite{Harmati2018, Harmati2019, Harmati2020}. Further, the behavior of FGCM models with sigmoid functions was analyzed, and sufficient conditions for the existence and uniqueness of fixed points were proposed \cite{Concepcion2020}.

The aforementioned studies address the convergence conditions of FCMs and FGCMs, their applications under various convergence states, and the development of learning algorithms leveraging FCM convergence. However, research on FGGCMs' convergence remains limited. This work seeks to bridge this gap by utilizing the Banach fixed point theorem and the Browder-Gohde-Kirk fixed point theorem to derive sufficient conditions for the convergence of FGGCMs to fixed points.

The contributions of this paper are as follows: 
\begin{itemize}
  \item This study establishes new sufficient conditions for the convergence of kernels and greyness in FGGCMs under the sigmoid activation function, addressing a gap in existing research.
  \item The proposed theorems are validated through various scenarios, offering a theoretical basis for the design, analysis, and development of learning algorithms for FGGCMs.
\end{itemize}

The organization of this paper is as follows:  
Section \ref{sec:Problem} provides an overview of the foundational definitions of FCM, FGCM, and FGGCM, along with the convergence theorems for FCM and FGCM under the sigmoid activation function. 
Section \ref{sec:convergence} establishes and proves sufficient conditions for FGGCMs to converge to fixed points when the sigmoid function is employed as the activation function.  
Section \ref{sec:result} presents designed examples to validate the proposed convergence theorems and discusses the results derived from these examples.  
Section \ref{sec:discussion} offers a detailed analysis and discussion of the convergence results.  
Finally, Section \ref{sec:conclusion} concludes the paper by summarizing the findings and proposing potential directions for future research on the convergence properties of FGGCMs.

\section{Preparatory Knowledge} \label{sec:Problem}
This section provides the foundational concepts required for the study. It begins with an explanation of the definition and reasoning process of FCM, followed by a presentation of the convergence theorems for FCM and FGCM under the sigmoid activation function. Finally, the reasoning framework of FGGCM is introduced, along with the necessary mathematical definitions to clarify the research problem addressed in this paper.

A FCM is formally represented as a 4-tuple \( M = \{\bm{C}, \mathbf{W}, \bm{A}, f\} \), where \( \bm{C} \) represents the set of all neurons in the map, \( \mathbf{W}: (C_i, C_j) \rightarrow w_{ij} \) specifies the causal weight square matrix, \( \bm{A}: C_i \rightarrow A^{(t)}_i \) denotes the function that determines the activation level of neuron \( C_i \) at discrete time step \( t \) (\( t = 1, 2, \dots, T \)), and \( f(\cdot) \) is the activation function applied to update the neurons' states.

The interaction between the components of an FCM is captured by Eq. \eqref{fcm_iter}, which outlines the iterative calculation of the state vector \( \bm{A^{(t)}} = [A^{(t)}_1, A^{(t)}_2, \cdots , A^{(t)}_M]^{\mathsf{T}} \) based on the initial state vector \( \bm{A^{(0)}} = [A^{(0)}_1, A^{(0)}_2, \cdots, A^{(0)}_M]^{\mathsf{T}} \):

\begin{equation}
  \label{fcm_iter}
  \bm{A^{( t+1 )}}= f \left(\mathbf{W} \bm{A^{(t )}} \right), 
\end{equation}

The elements of the weight matrix are typically bounded within the range \([-1, 1]\). In some cases, recursive interactions are restricted to prevent implicit self-reinforcing causal effects, achieved by setting the diagonal elements of the weight matrix to zero (\( w_{ii} = 0 \)), ensuring that variables do not influence their own states. Node activation values, denoted as \( \bm A \), are generally confined to the intervals \([0, 1]\) or \([-1, 1]\). For discrete activation states, binary and trivalent functions are commonly applied as activation functions. Conversely, in continuous activation states, the sigmoid function (Eq. \ref{sigmoid}) or the hyperbolic tangent function is often used.

\begin{equation}
  \label{sigmoid}
  S(x) = \frac{1}{1 + \me^{-\lambda x}}
\end{equation}

In Eq. (\ref{sigmoid}), the parameter \(\lambda\) serves to regulate the slope of the activation function, with \(\lambda > 0\). The value of \(\lambda\) significantly influences the convergence characteristics of the FCM, as it determines the sensitivity of the activation function to changes in input.
As the iterative reasoning of FCMs advances, the activation states of their nodes can converge to fixed points, limit cycles, or chaotic states. These behaviors are formally defined as follows:

\begin{defi}
  Fixed point: $\exists t_\alpha \in \{1, 2, \ldots, T - 1\}, \forall t \geq t_\alpha, s.t. \bm A^{(t+1)} = \bm A^{(t)}$. The system enters a state of equilibrium after $t_\alpha$ , resulting in a consistent output where $\bm A^{(t_\alpha) }= \bm A^{(t_\alpha+1)} = \bm A^{(t_\alpha+2)} = \ldots = \bm A^{(T)}$.
\end{defi}

\begin{defi}
  Limit cycle: $\exists t_\alpha, P \in \{1, 2, \cdots, T - 1\}, \forall t \geq t_\alpha, s.t. \bm A^{(t+P)} = A^{(t)}$. The system enters a periodic state after $t_\alpha$ step, resulting in a repeated output pattern such that $\bm A^{(t_\alpha)} = \bm A^{(t_\alpha+P)} = \bm A^{(t_\alpha+2P) }= \ldots = \bm A^{(t_\alpha+jP)}$, where $j \in \mathbb{N}$.
\end{defi}

\begin{defi}
  Chaos: The system persistently generates distinct state vectors over consecutive steps. In these instances, the FCM fails to reach a stable state, resulting in unpredictable and ambiguous system responses.
\end{defi}

For simplicity, references to the convergence, kernel, and greyness of FCM, FGCM, and FGGCM in this paper pertain to the convergence of node activation values, the kernel of node activation values, and the greyness of node activation values within these models.

\begin{figure}[htbp]
  \centering
  \begin{tikzpicture}[every node/.style={circle,draw,inner sep=5pt,font=\footnotesize}]
    \def\side{2}
  \foreach \i in {1,2,...,7}  {        \pgfmathsetmacro{\angle}{\i * 51.42857142857143} 
  \pgfmathsetmacro{\x}{\side * cos(\angle)}
  \pgfmathsetmacro{\y}{\side * sin(\angle)}
  \node[fill=blue!20] (node\i)at (\x,\y){$C_{\i}$};};
  \foreach \i in{1,2,...,7} {\foreach \j in{1,2,...,7}  \ifnum \i = \j
  \else \draw[-latex] (node\i)--(node\j)
\fi;} 
  \draw[-latex] (node3.north) .. controls (-2,2) and (-2.2,1.5 ) .. (node3.north west); 
  \draw[-latex] (node4.west) .. controls (-3,-2) and (-2,-1.5 ) .. (node4.south); 
  \draw[-latex] (node5.south west) .. controls (-1,-3) and (-0.5,-3 ) .. (node5.south);
  \draw[-latex] (node6.south) .. controls (1,-3) and (2,-2.5 ) .. (node6.south east);
  \draw[-latex] (node7.south east) .. controls (2.5,-1) and (3.5,0 ) .. (node7.east);
  \end{tikzpicture}
  \caption{The Web Experience FCM constructed for users.}
  \label{figfcmexample}
\end{figure}
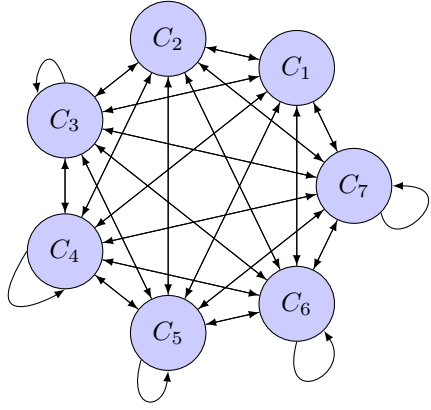

For instance, Fig. \ref{figfcmexample} demonstrates the use of FCMs to model a user's web experience on the World Wide Web \cite{Meghabghab2001,Meghabghab2003}, with the corresponding meanings of the nodes provided in Table \ref{nodes_web}.
For convenience, this FCM will hereafter be referred to as the Web Experience FCM.

\begin{table}[htbp]
  \caption{The interpretation of the nodes in the Web Experience FCM.}\label{nodes_web}
  \centering
  \begin{tabular*}{.35\linewidth}{cc}
    \toprule    
    Nodes & Meanings \\ 
    \midrule 
    $C_1 $& Exploration  \\ 
    $C_2 $& Scanning   \\ 
    $C_3 $& Temporal Restrictions  \\ 
    $C_4 $& Data Restrictions  \\ 
    $C_5 $& Achievement  \\ 
    $C_6 $& Pertinence \\ 
    $C_7 $& Unsuccess  \\ 
    \bottomrule    
    \end{tabular*}
  \end{table}

This FCM uses the sigmoid function as its activation function, 
and its corresponding weight matrix is as follows:
\begin{equation}
  \label{wweb}
  \setlength{\arraycolsep}{2pt}
\begin{split}
    &\mathbf{W}=\\
  &\begin{pmatrix}
    0.0 & -0.9 & -0.88 & 1.0 & -0.85 & -0.83 & 1.0 \\
    1.0 & 0.0 & -0.93 & -0.89 & -0.9 & -0.94 & 1.0 \\
    -0.98 & -0.93 & -1.0 & -1.0 & 1.0 & 1.0 & 1.0 \\
    -0.99 & -0.89 & -1.0 & -0.39 & 0.73 & 0.58 & 0.7 \\
    1.0 & 1.0 & 1.0 & 1.0 & -0.8 & 0.51 & 1.0 \\
    1.0 & 1.0 & 0.83 & 1.0 & 0.51 & -0.39 & 1.0 \\
    1.0 & 1.0 & 1.0 & 1.0 & -0.71 & -0.49 & -0.67
    \end{pmatrix}
\end{split}
\end{equation}

\begin{figure}[htbp]
  \centering
    \includegraphics[width=0.6\linewidth]{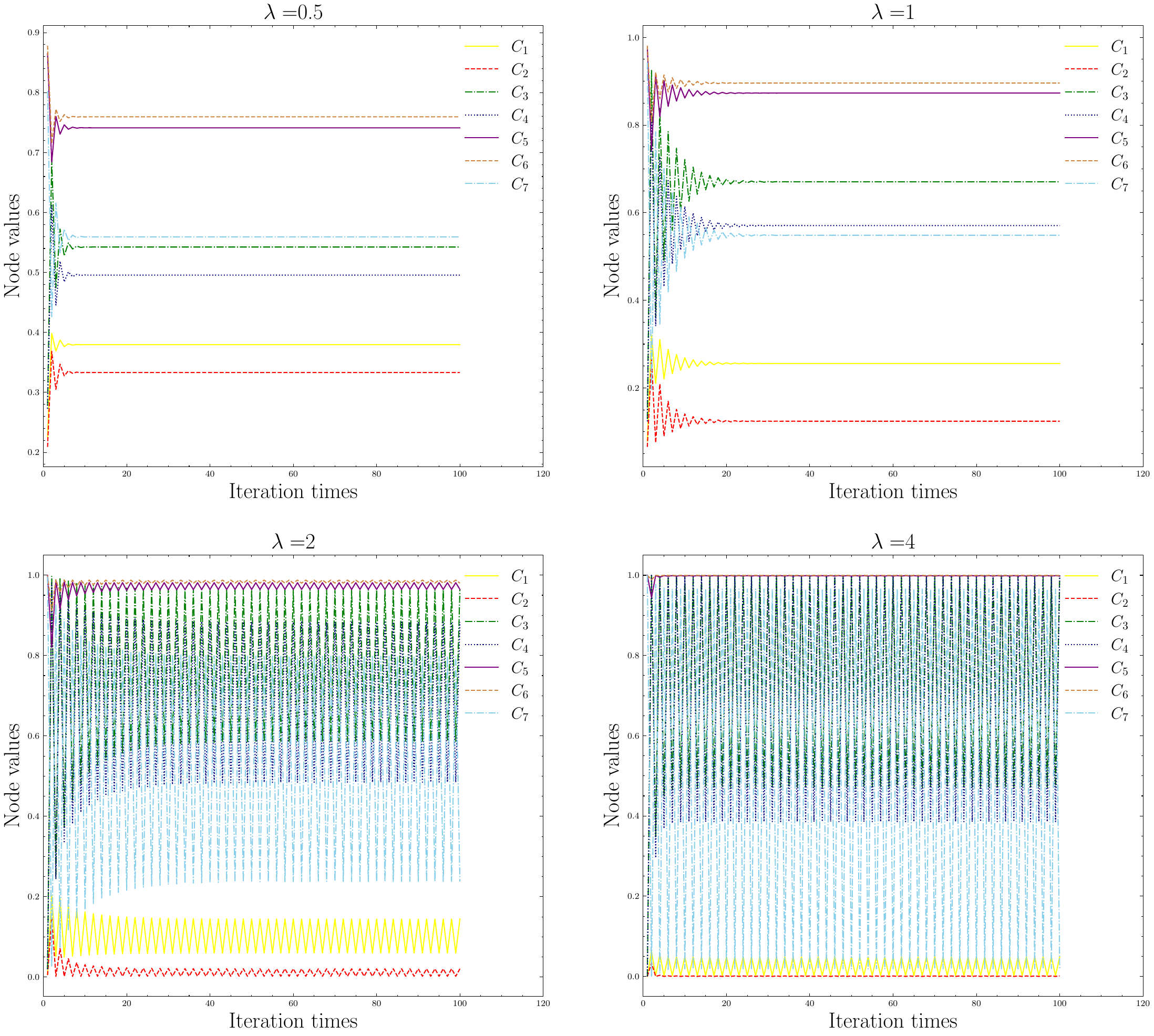}
  \caption{The output results of the Web Experience FCM across varying \(\lambda\) values.}
  \label{figfcmweb}
\end{figure}

  Starting with an initial vector of \(\setlength{\arraycolsep}{2pt}
\begin{pmatrix}
  1 & 1 & 1 & 1 & 1 & 1 & 0
  \end{pmatrix}\), Fig. \ref{figfcmweb} shows the simulation results for each node with \(\lambda\) values set to $0.5$, $1$, $2$, and $4$.

As illustrated in Fig. \ref{figfcmweb}, the FCM demonstrates convergence to a fixed point for \(\lambda\) values of $0.5$ and $1$, whereas for \(\lambda\) values of $2$ and $4$, the system exhibits a limit cycle behavior.

The conditions ensuring the existence of fixed points for an FCM employing a sigmoid activation function can be summarized as follows:

\begin{thm}
  \label{thmfcm}
  In an FCM employing the sigmoid activation function \( f(x) = \frac{1}{1 + \me^{-\lambda x}} \), a unique fixed point exists for each concept value \( A_i \) if the following condition is satisfied:  
  \begin{equation}
    \label{eqfcmsig}
    \left\lVert \mathbf{W} \right\rVert _F  < \frac{4}{\lambda}.
  \end{equation} 
When the condition \(\left\lVert \mathbf{W} \right\rVert _F = \frac{4}{\lambda}\) holds, at least one fixed point is guaranteed for each concept value \( A_i \).
The term \( \left\lVert \mathbf{W} \right\rVert _F \) represents the Frobenius norm of the weight matrix \( \mathbf{W} \). It is computed as:
\[
\| \mathbf{W} \|_{F} = \left( \sum_{i=1}^{n} \sum_{j=1}^{n} w_{ij}^2 \right)^{\frac{1}{2}},
\]
where \( w_{ij} \) denotes the individual elements of the matrix \( \mathbf{W} \), and \( n \) is the dimension of the square matrix.
\end{thm}

The Theorem \ref{thmfcm} provides a sufficient condition for the convergence of FCM to a fixed point, addressing a gap in the understanding of FCM convergence. However, due to limitations in data formatting, this theorem cannot be applied to FGCMs that use IGNs or FGGCMs that incorporate GGNs. The criterion for the existence of fixed points in FGCM is as follows:
\begin{thm}
  \label{thmfgcm}
If \( \otimes \mathbf{W} \) represents the weight matrix of a FGCM, which includes potential feedback, where the weights \( \otimes w_{ij} = \left[\underline{w_{ij}}, \overline{w_{ij}}\right] \) are interval grey numbers that are either nonnegative or nonpositive, then the elements of matrix \( \mathbf{W}^* \) are defined as follows:
\begin{equation}
  \label{wstar}
  w_{i j}^{*}=\left\{\begin{array} {c c} {{ \left  \vert \underline{w_{ij}} \right \vert}} & \mathrm{if} \,\, \underline{w_{ij}}\leqslant \overline{w_{ij}} \leqslant 0 \\ 
    \overline{w_{ij}}& \mathrm{if} \,\, 0 \leqslant \underline{w_{ij}}\leqslant \overline{w_{ij}}   \\ \end{array} \right.
\end{equation}
Let \( \lambda > 0 \) be the parameter of the sigmoid function used in the iterative process. If the following condition holds:
\[
\left\lVert \mathbf{W}^* \right\rVert _F < \frac{4}{\lambda},
\]
then the FGCM will converge to a unique grey fixed point, irrespective of the initial concept values.

\end{thm}

The Theorem \ref{thmfgcm} presents the convergence conditions for FGCM when the elements of its weight matrix are either entirely positive or entirely negative, filling a gap in the convergence analysis of FGCM. However, it cannot handle cases where the weight matrix contains elements of the form \([a, -b]\), where \(a, b > 0\) and \(a, b \in \mathbb{R}\), as in these situations the \(\mathbf{W}^*\) matrix cannot be computed. Additionally, due to limitations in data formatting, this theorem is not applicable to the convergence analysis of FGGCM.

The FGGCM extends the capabilities of the FGCM, with the primary aim of improving its ability to model uncertainty. This enhancement is accomplished by integrating the more comprehensive uncertainty-handling features offered by Grey System Theory, which FGCM only partially exploits. Unlike FGCM, which is based on interval grey numbers (IGN), FGGCM utilizes general grey numbers (GGN) as its foundational element. A GGN is a broader form of an IGN, defined as \( g^\pm \in \bigcup\limits^n_{i=1} [\underline{a}_i, \overline{a}_i] \). For example, \( [0, 1.2] \cup [1.5, 2] \cup [3, 5] \cup 6 \) constitutes a GGN, whereas \( [0, 1] \) serves as both an IGN and a special case of a GGN. The mathematical operations involving GGNs and IGNs differ: while IGNs rely on their upper and lower bounds for calculations \cite{Qiao2024}, GGNs use their kernel and greyness. Both types of grey numbers have a kernel, which represents the most probable crisp value within the grey number's range. It is precisely the differences in the computational methods of GGN and IGN that result in distinct forms of activation functions, as well as variations in the reasoning process of FGCM and FGGCM, ultimately leading to differences in their respective convergence conditions.

The greyness value quantifies the uncertainty level of a GGN \( g^\pm \). For simplicity, in this text, the greyness degree will be referred to as greyness. A GGN can be expressed in terms of its kernel and greyness using the notation \( \hat{g}_{g^\circ} \), where \( \hat{g} \) represents the kernel and \( g^\circ \) denotes the greyness. This representation allows a GGN to be written as \( g^\pm = \hat{g}_{g^\circ} \) or \( \otimes g = \hat{g}_{g^\circ} \), whereas an IGN is typically expressed as \( \otimes g = [\underline{g}, \overline{g}] \). Comprehensive details regarding the operational rules of IGNs and the reasoning framework of FGCM can be found in \cite{Salmeron2010, Chen2021, Salmeron2019}. Similarly, in-depth discussions on the operations of GGNs and the inference mechanisms of FGGCM are provided in \cite{Chen2020, Chen2021}.

The sigmoid function in the form of GGN, as referenced in \cite{Chen2020}, is defined as follows:
\begin{equation}
  \label{ggsigmoid}
  S(g^\pm)=\frac{1}{1+\me^{-\lambda g^\pm}}.
\end{equation}
$S(g^\pm)$ is a new GGN, the kernel is
\begin{equation}
  \hat {S}(g^\pm)=\frac{1}{1+\me^{-\lambda \hat {g}}},
  \label{sk}
\end{equation}
the greyness is
\begin{equation}
  S^\circ(g^\pm)= \frac{1}{1+\me^{-\lambda \hat{g}}}g^\circ.
  \label{gsigmoid}
\end{equation} 

Similar to Eq. \eqref{sigmoid}, the parameter $\lambda$ in Eqs. \eqref{ggsigmoid} to \eqref{gsigmoid} is a positive constant that determines the steepness of the corresponding curve or surface.

\section{The Convergence of sigmoid FGGCM} \label{sec:convergence}

This section focuses on establishing and proving the convergence conditions of FGGCM with the sigmoid function as the activation function. First, three lemmas essential for proving the convergence conditions of FGGCM are presented. Next, the convergence conditions and their proofs for FGGCM under the sigmoid activation function are provided. Finally, several related corollaries are discussed.

The most common approach for determining fixed points involves iterative methods. Starting with an arbitrary initial point \( x_0 \in X \), the sequence is defined as \( x_1 = T x_0 \), \( x_2 = T x_1 = T^2 x_0 \), and so on, where \( T \) is a contraction mapping. This iterative process is similar to the reasoning mechanism in FCM and FGGCM models. Consequently, investigating the convergence conditions of an FGGCM is equivalent to determining the conditions under which the FGGCM is a contraction mapping. For contractive mappings and non-expansive mappings, there are the Banach Fixed Point Theorem and the Browder-Gohde-Kirk Fixed Point Theorem, as stated in Lemmas \ref{Banach} and \ref{Browder}.

\begin{lem}[Banach Fixed Point Theorem]
  \label{Banach}

Let \( X \) be a complete metric space and \( T: X \rightarrow X \) a mapping that satisfies the contraction condition:  
\[ d(Tx, Ty) \leqslant k \cdot d(x, y), \]  
for all \( x, y \in X \), where \( 0 \leqslant k < 1 \). Such a mapping \( T \) is termed a contraction mapping.  

Under these conditions, there exists a unique element \( x^* \in X \) such that:  
\[ T(x^*) = x^*. \]  
This unique element \( x^* \) is referred to as the fixed point of the mapping \( T \).
\end{lem}

\begin{lem}[Browder-Gohde-Kirk Fixed Point Theorem]
  \label{Browder}
  Let \( X \) be a Banach space, and \( C \) a subset of \( X \) that is closed, bounded, and convex. A mapping \( U: C \to C \) is called a nonexpansive mapping if, for any \( x, y \in C \), the following inequality is satisfied:  
\[ \| Ux - Uy \| \leqslant \| x - y \|, \]  
where \( \| \cdot \| \) denotes the norm in \( C \).  

If \( X \) is a uniformly convex Banach space and \( U \) is a nonexpansive mapping on the closed, bounded, and convex subset \( C \), then \( U \) is guaranteed to have at least one fixed point in \( C \).
\end{lem}

The proof of the convergence conditions for FGGCM also relies on the range of values of the derivative of the sigmoid function, as stated in Lemma \ref{sig_bound}.
\begin{lem}
  \label{sig_bound}

Given the sigmoid function \( f(x) = \mathrm{sigmoid}(x) = \frac{1}{1 + \me^{-\lambda x}}\), where \( x \in \mathbb{R} \) and \( \lambda > 0 \), it can be shown that for any two real numbers \( a, b \in \mathbb{R} \), the following inequality holds:
\[
|f(b) - f(a)| \leqslant \frac{\lambda}{4} |b - a|.
\]

\end{lem}
\begin{pf}
  By differentiating $f(x)$, it can be obtained:
\begin{equation}
  \label{sigmoidderivative}
  f'(x) = \frac{\lambda}{2 + \me^{-\lambda x} + \frac{1}{\me^{-\lambda x}}} \leqslant \frac{\lambda}{4}.
\end{equation}
The equality holds if and only if $x = 0$. According to the Lagrange Mean Value Theorem, if a function $f(x)$ is continuous on the closed interval $[a, b]$ and differentiable on the open interval $(a, b)$, then there exists at least one point $\xi$ in the open interval $(a, b)$ such that:
$$
f(b) - f(a) = f'(\xi)(b - a)
$$
Substituting Eq. \eqref{sigmoidderivative} into the Lagrange Mean Value Theorem, we obtain:
\begin{equation*}
  |f(b) - f(a)| \leqslant \frac{\lambda}{4}|(b - a)|
\end{equation*}

\end{pf}

The following part presents the convergence condition for FGGCM with the sigmoid function as the activation mechanism, along with the associated proof.
\begin{thm}
  \label{thmFGGCMsig}
Given that \( \mathbf{W^\pm} = \mathbf{\hat{W}_{W^\circ}} \) represents the weight matrix of the FGGCM, where $w^\pm_{ij} = \hat{w}{_{ij}}_{w_{ij}^\circ} \in \mathbb{G}$ are its elements, \(\mathbb{G}\) represents the set composed of all GGNs, and \( \lambda > 0 \) is the parameter of the sigmoid activation function for the FGGCM model with \( n \) nodes, the sigmoid function is defined as:
\[
  \mathrm{sigmoid}(g^\pm) = \frac{1}{1+\me^{-\lambda g^\pm}}.
\]
The FGGCM will converge to a unique fixed point if the following conditions hold:
\begin{equation}
  \label{sigkernelconvergence}
  \left( \sum_{i=1}^n \sum_{j=1}^n \hat{w}_{ij}^2 \right)^{\frac{1}{2}} < \frac{4}{\lambda}
\end{equation}
and
\begin{equation}
  \label{siggreynessconvergence}
  \left( \sum_{i=1}^{n} \sum_{j=1}^{n} \left( \frac{\hat{A}_i \left| \hat{A}_j \hat{w}_{ij} \right| \theta \left( C_j^\circ - \hat{w}_{ij}^\circ \right)} {\sum_{j=1}^{n} \left| \hat{w}_{ij} \hat{A}_j \right|} \right)^2 \right)^{\frac{1}{2}} < 1.
\end{equation}
Here,\(A_i,A_j\) represents the value of $i$ and $j$ node of FGGCM. \(C_j^\circ\) represents the greyness of node \(C_j\) in the FGGCM at any given iteration step. The function \(\theta(x)\) is the Heaviside step function, defined as:
\[
\theta(x) = \begin{cases}
0 & \text{if } x < 0, \\
1 & \text{if } x \geq 0.
\end{cases}
\]
In particular, if the following conditions hold:
\begin{equation}
  \label{sigeqkernel}
  \left( \sum_{i=1}^n \sum_{j=1}^n \hat{w}_{ij}^2 \right)^{\frac{1}{2}} = \frac{4}{\lambda},
\end{equation}
and
\begin{equation}
  \label{sigeqgreyness}
  \left( \sum_{i=1}^{n} \sum_{j=1}^{n} \left( \frac{\hat{A}_i \left| \hat{A}_j \hat{w}_{ij} \right| \theta \left( C_j^\circ - \hat{w}_{ij}^\circ \right)} {\sum_{j=1}^{n} \left| \hat{w}_{ij} \hat{A}_j \right|} \right)^2 \right)^{\frac{1}{2}} = 1,
\end{equation}
then the FGGCM will have at least one fixed point.
\end{thm}

\begin{pf}
  \(\mathbb{G}^n\) denotes the set consisting of all GGN vectors. $\forall \bm{A,A'} \in \mathbb{G}^n$, define the metric $d$ on $\mathbb{G}^n$ as follows:
  \begin{equation}
      d(\bm{A^\pm, {A'}^\pm}) = \left(\sum_{i=1}^{n} \left( \left|\hat{A_i} - \hat{{A'}_i} \right|^2 + \left| A_i^\circ - {A'}_i^\circ \right|^2 \right) \right)^{\frac{1}{2}}.
  \end{equation}

It is straightforward to demonstrate that \(d\) satisfies symmetry, non-negativity, non-degeneracy, and the triangle inequality. The triangle inequality, in particular, can be derived from the Minkowski inequality.

Calculate $d^2(f(\mathbf{W^\pm }\bm A),f(\mathbf{W^\pm} \bm A'))$ as Eq. \eqref{dSAA},

  \begin{equation}
    \label{dSAA}
  \begin{split}
    &d^2(f(\mathbf{W^\pm }\bm A),f(\mathbf{W^\pm} \bm A'))\\ 
    =& \sum_{i=1}^n \left(\frac{1}{1+\me^{-\lambda \bm{\hat{W_i} \hat{A}}}} - \frac{1}{1+\me^{-\lambda \bm{\hat{W_i} \hat{A'}}}} \right)^2 + GRS\\
    \leqslant &  \sum_{i=1}^n \left(\frac{\lambda}{4}\right)^2 \left(\bm{\hat{W_i} \hat{A}} - \bm{\hat{W_i} \hat{A'}} \right)^2 + GRS\\
    = &\left(\frac{\lambda}{4}\right)^2  \sum_{i=1}^n \left(\sum_{j=1}^n \hat{w}_{ij}\left(\hat{A}_j - \hat{A'}_j\right)\right)^2 + GRS\\
    \leqslant & \left(\frac{\lambda}{4}\right)^2 \sum_{i=1}^n \left(\sum_{j=1}^n \hat{w}_{ij}^2 \right) \left(\sum_{j=1}^n \left(\hat{A}_j - \hat{A'}_j\right)^2 \right)+ GRS\\
  \end{split}
  \end{equation}

where 
\begin{equation*}
  \begin{split}
    GRS= &\sum_{i=1}^n \left(\left(\frac{1}{1+\me^{-\lambda \bm{\hat{W_i} \hat{A}}}}\right)^\intercal\bm{(W_iA)^\circ}-\right.\\
    &\left.\left(\frac{1}{1+\me^{-\lambda \bm{\hat{W_i} \hat{A'}}}}\right)^\intercal\bm{(W_iA')^\circ}\right)^2
  \end{split}
\end{equation*}

In Eq. (\ref{dSAA}), the first inequality is obtained using Lemma \ref{sig_bound}, and the second inequality follows directly from the application of the Cauchy-Schwarz inequality.

On the other hand,
\begin{equation}
  \label{dAA}
    d^2(\bm{A, A'}) = \sum_{j=1}^n \left(\hat{A}_j - \hat{A'}_j\right)^2 + \sum_{j=1}^n \left(A_j^\circ - {A'}_j^\circ\right)^2
\end{equation}

By examining Eqs. (\ref{dSAA}) and (\ref{dAA}), it can be observed that the distance is split into two components: the kernel and the greyness. Notably, the convergence of the greyness does not influence the convergence of the kernel.

By focusing on the kernel component, it can be inferred that if the following inequality holds:  
\begin{equation*}
  \begin{split}
    \left(\frac{\lambda}{4}\right)^2 \sum_{i=1}^n \left(\sum_{j=1}^n \hat{w}_{ij}^2 \right) \left(\sum_{j=1}^n \left(\hat{A}_j - \hat{A'}_j\right)^2 \right) <\sum_{j=1}^n \left(\hat{A}_j - \hat{A'}_j\right)^2
  \end{split}
\end{equation*}
this condition is equivalent to satisfying the inequality:  
\begin{equation}
  \label{sigkernelconvergenceinproof}
  \left(\sum_{i=1}^n\sum_{j=1}^n \hat{w}_{ij}^2 \right)^\frac{1}{2} < \frac{4}{\lambda}.
\end{equation}  
Under this condition, the kernel of the FGGCM is guaranteed to converge to a unique fixed point.

When employing the sigmoid function as the activation mechanism, the iterative process governing the greyness can be formulated as follows:
\begin{equation}
  \label{sig_grey_iter}
  \begin{split}
    {A^\circ_j}^{t + 1}=&f({A_1^\circ}^t,{A_2^\circ}^t,\cdots,{A_n^\circ}^t,\hat{A}_1^t,\hat{A}_2^t\cdots \hat{A}_n^t)\\
     = &\frac{1}{\left(1+\me^{-\lambda \sum_{j=1}^{n}\hat{w}_{ij}\hat{A}_j^t}\right)}\sum_{j=1}^{n}\frac{\max (w_{ij}^\circ,{A_j^\circ}^t)\left|\hat{w}_{ij}\hat{A}_j^t\right|}{\sum_{j=1}^{n}\left|\hat{w}_{ij}\hat{A}_j^t\right|}\\
     = &\hat{A}_i^{t+1}\sum_{j=1}^{n}\frac{ \max (w_{ij}^\circ,{A_j^\circ}^t)\left|\hat{w}_{ij}\hat{A}_j^t\right|}{\sum_{j=1}^{n}\left|\hat{w}_{ij}\hat{A}_j^t\right|}.\\
  \end{split}
\end{equation}

Based on the mean value theorem for multivariable functions, Eq. \eqref{multivariate} is satisfied. 
\begin{equation}
  \label{multivariate}
  \begin{split}
    &f({A_1^\circ},{A_2^\circ},\cdots,{A_n^\circ},\hat{A}_1,\hat{A}_2\cdots \hat{A}_n) \\
    &-  f({{A'}_1^\circ},{{A'}_2^\circ},\cdots,{{A'}_n^\circ},\hat{{A'}}_1,\hat{{A'}}_2\cdots \hat{{A'}}_n)\\
    =& \bigtriangledown f \cdot \bigtriangleup  \bm{A}\\
    =& \begin{pmatrix}
      \frac{\partial f}{\partial C_1^\circ} &\frac{\partial f}{\partial C_2^\circ} & \cdots&\frac{\partial f}{\partial C_n^\circ} &\frac{\partial f}{\partial \hat{C}_1}&\frac{\partial f}{\partial \hat{C}_2}&\cdots &\frac{\partial f}{\partial \hat{C}_n}\\
    \end{pmatrix} \cdot \\
    &\begin{matrix}
      ({A_1^\circ} - {{A'}_1^\circ} &{A_2^\circ} - {{A'}_2^\circ}& \cdots&{A_3^\circ} - {{A'}_3^\circ}\\
      \hat{A}_1 - \hat{{A'}}_1&\hat{A}_2 - \hat{{A'}}_2&\cdots &\hat{A}_n - \hat{{A'}}_n ) ^\top\\
    \end{matrix}\\
    =& \sum_{j=1}^{n} \frac{\partial f}{\partial C_j^\circ} ({A_j^\circ} - {{A'}_j^\circ}) + \sum_{j=1}^{n} \frac{\partial f}{\partial \hat{C}_j} (\hat{A}_j - \hat{{A'}}_j)
  \end{split}
\end{equation}
where, $\hat{C}_j = \hat{A}_j +t(\hat{{A'}}_j-\hat{A}_j)$, $C_j^\circ = {{A}_j^\circ}+t({{A'}_j^\circ} - {{A}_j^\circ})$,$0<t<1$.
It can be found that in the distance measurement between greyness, the kernel contributes a part of the distance.
Subsequently, the partial derivative with respect to the greyness is computed as follows:
\begin{equation}
  \label{dsg}
  \begin{split}
    \frac{\partial f}{\partial C_j^\circ}= &\frac{\left|{\hat{C}_{j} w_{ij}}\right| \theta\left(C_j^\circ - \hat{w}_{ij}^\circ\right)}{\left(1+\me ^{\sum_{j=1}^{n}\hat{w}_{ij}\hat{C}_j} \right)\sum_{j=1}^{n}\left|\hat{w}_{ij}\hat{C}_j\right|}\\
    =&\frac{\hat{C}_i^{t+1} \left|{\hat{C}_{j} w_{ij}}\right| \theta\left(C_j^\circ - \hat{w}_{ij}^\circ\right)}{\sum_{j=1}^{n}\left|\hat{w}_{ij}\hat{C}_j\right|}\\
  \end{split}
\end{equation}

Assuming that the kernel of the FGGCM has already converged to a fixed point, the impact of the kernel on the greyness can be disregarded. Consequently, the computation of the greyness component in Eq. \eqref{dSAA} is expressed as Eq. \eqref{sig_grey_cal}:

  \begin{equation}
    \label{sig_grey_cal}
    \small
    \begin{split} 
      &\sum_{i=1}^n \left(\sum_{j=1}^{n}\frac{\max(w_{ij}^\circ,A_j^\circ)\left|\hat{w}_{ij}\hat{A}_j\right|}{\left(1+\me^{-\lambda \sum_{j=1}^{n}\hat{w}_{ij}\hat{A}_j}\right)\sum_{j=1}^{n}\left|\hat{w}_{ij}\hat{A}_j\right|}-  \sum_{j=1}^{n}\frac{\max(w_{ij}^\circ,{A'}_j^\circ)\left|\hat{w}_{ij}\hat{A'}_j\right|}{\left(1+\me^{-\lambda \sum_{j=1}^{n}\hat{w}_{ij}\hat{A'}_j}\right)\sum_{j=1}^{n}\left|\hat{w}_{ij}\hat{A'}_j\right|}\right)^2\\ 
      =&\sum_{i=1}^{n}\left(f({A_1^\circ},{A_2^\circ},\cdots,{A_n^\circ},\hat{A}_1,\hat{A}_2\cdots \hat{A}_n) -   f({{A'}_1^\circ},{{A'}_2^\circ},\cdots,{{A'}_n^\circ},\hat{{A'}}_1,\hat{{A'}}_2\cdots \hat{{A'}}_n)\right)^2\\
      =& \sum_{i=1}^{n} \left(\sum_{j=1}^{n}\frac{\partial f}{\partial C_j^\circ} ({A_j^\circ} - {{A'}_j^\circ})+ \sum_{j=1}^{n}\frac{\partial f}{\partial \hat{C}_j} ({\hat{A}_j} - {\hat{A'}_j})\right)^2\\
      =& \sum_{i=1}^{n} \left(\sum_{j=1}^{n}\frac{\left|{\hat{C}_{j} w_{ij}}\right| \theta\left(C_j^\circ - \hat{w}_{ij}^\circ\right)}{\left(1+\me ^{\sum_{j=1}^{n}\hat{w}_{ij}\hat{C}_j} \right)\sum_{j=1}^{n}\left|\hat{w}_{ij}\hat{C}_j\right|} ({A_j^\circ} - {{A'}_j^\circ})\right)^2\\
      \leqslant & \sum_{i=1}^{n} \sum_{j=1}^{n}\left(\frac{\left|{\hat{C}_{j} w_{ij}}\right| \theta\left(C_j^\circ - \hat{w}_{ij}^\circ\right)}{\left(1+\me ^{\sum_{j=1}^{n}\hat{w}_{ij}\hat{C}_j} \right)\sum_{j=1}^{n}\left|\hat{w}_{ij}\hat{C}_j\right|}\right)^2\sum_{j=1}^{n} \left({A_j^\circ} - {{A'}_j^\circ}\right)^2 \\
    \end{split}
  \end{equation}

Given that \(\hat{A}_j = \hat{{A'}}_j\), implying \(\hat{C}_j = \hat{A}_j + t(\hat{{A'}}_j - \hat{A}_j) = \hat{A}_j\), the following conditions ensure that the greyness of the sigmoid-based FGGCM will necessarily converge to a unique fixed point:
 
If  
\[
\sum_{i=1}^{n} \sum_{j=1}^{n}\left(\frac{\left|{\hat{A}_{j} \hat{w}_{ij}}\right|\theta\left(C_j^\circ - \hat{w}_{ij}^\circ\right) }{\left(1+\me ^{-\lambda \sum_{j=1}^{n}\hat{w}_{ij}\hat{A}_j} \right)\sum_{j=1}^{n}\left|\hat{w}_{ij}\hat{A}_j\right|}\right)^2 < 1,
\]
or equivalently, if  
\[
\left(\sum_{i=1}^{n} \sum_{j=1}^{n}\left(\frac{\hat{A}_{i}\left|{\hat{A}_{j} \hat{w}_{ij}}\right|\theta\left(C_j^\circ - \hat{w}_{ij}^\circ\right) }{\sum_{j=1}^{n}\left|\hat{w}_{ij}\hat{A}_j\right|}\right)^2\right)^\frac{1}{2} < 1,
\]
then the greyness of the sigmoid FGGCM will converge to a unique fixed point.

By applying Lemma \ref{Browder}, it follows that if both Eq. \eqref{sigeqkernel} and Eq. \eqref{sigeqgreyness} hold simultaneously, the FGGCM is guaranteed to possess at least one fixed point.

\end{pf}

 The following corollaries can be straightforwardly deduced from the preceding proof process.

\begin{corollary}
  If Eq. \eqref{sigkernelconvergence} is satisfied, the kernel of each node in the sigmoid-based FGGCM is guaranteed to converge to a unique fixed point. Conversely, if Eq. \eqref{sigeqkernel} is fulfilled, the kernel of the FGGCM is assured to possess at least one fixed point.
\end{corollary}

\begin{corollary}
  If the kernel of the sigmoid-based FGGCM converges and inequality \eqref{siggreynessconvergence} is satisfied, the greyness of each node in the FGGCM is guaranteed to converge to a unique fixed point. Conversely, if the kernel of the FGGCM converges and Eq. \eqref{sigeqgreyness} is met, the greyness of each node in the FGGCM is ensured to have at least one fixed point.
\end{corollary}

Through an analysis of the greyness iteration process in the FGGCM, Corollary \ref{sig_grey_convergence} is derived. Notably, Eq. \eqref{sig_grey_convergence_eq} represents a particular case of the general condition presented in Eq. \eqref{siggreynessconvergence}, as stated in Theorem \ref{thmFGGCMsig}. This insight further validates and strengthens the convergence criteria for greyness outlined in Theorem \ref{thmFGGCMsig}, specifically in Eq. \eqref{siggreynessconvergence}.

\begin{corollary}
  \label{sig_grey_convergence}
   If the kernel of an FGGCM employing the sigmoid activation function converges, and for any iteration step the condition \(\max(w_{ij}^\circ, A_j^\circ) = A_j^\circ\) holds, then the greyness of the FGGCM must converge to a unique fixed point under the condition:  
   \begin{equation} \label{sig_grey_convergence_eq}
    \sum_{i=1}^{n} \sum_{j=1}^{n}\left(\frac{\hat{A}_{i}\left|\hat{A}_{j} \hat{w}_{ij}\right|}{\sum_{j=1}^{n}\left|\hat{w}_{ij}\hat{A}_j\right|}\right)^2 = {\left\lVert \mathbf{M}\right\rVert }^2_F <1
  \end{equation}
This fixed point is the characteristic vector associated with the eigenvalue \(1\) of \(\mathbf{M}\).
   where
   \begin{equation}
    \label{sig_M}
    \mathbf{M} = 
    \begin{pmatrix}
      \frac{\hat{A}_1^{t+1}\left|\hat{w}_{11}\hat{A}_1^t\right|}{\sum_{j=1}^{n}\left|\hat{w}_{1j}\hat{A}_j^t\right|} & \frac{\hat{A}_1^{t+1}\left|\hat{w}_{12}\hat{A}_2^t\right|}{\sum_{j=1}^{n}\left|\hat{w}_{1j}\hat{A}_j^t\right|} & \cdots &  \frac{\hat{A}_1^{t+1}\left|\hat{w}_{1n}\hat{A}_n^t\right|}{\sum_{j=1}^{n}\left|\hat{w}_{1j}\hat{A}_j^t\right|}\\
      \frac{\hat{A}_2^{t+1}\left|\hat{w}_{21}\hat{A}_1^t\right|}{\sum_{j=1}^{n}\left|\hat{w}_{2j}\hat{A}_j^t\right|} & \frac{\hat{A}_2^{t+1}\left|\hat{w}_{22}\hat{A}_2^t\right|}{\sum_{j=1}^{n}\left|\hat{w}_{2j}\hat{A}_j^t\right|} & \cdots &  \frac{\hat{A}_2^{t+1}\left|\hat{w}_{2n}\hat{A}_n^t\right|}{\sum_{j=1}^{n}\left|\hat{w}_{2j}\hat{A}_j^t\right|}\\
      \vdots &\vdots & \ddots & \vdots \\
      \frac{\hat{A}_n^{t+1}\left|\hat{w}_{n1}\hat{A}_1^t\right|}{\sum_{j=1}^{n}\left|\hat{w}_{nj}\hat{A}_j^t\right|} & \frac{\hat{A}_n^{t+1}\left|\hat{w}_{n2}\hat{A}_2^t\right|}{\sum_{j=1}^{n}\left|\hat{w}_{nj}\hat{A}_j^t\right|} & \cdots &  \frac{\hat{A}_n^{t+1}\left|\hat{w}_{nn}\hat{A}_n^t\right|}{\sum_{j=1}^{n}\left|\hat{w}_{nj}\hat{A}_j^t\right|}\\
    \end{pmatrix}.
  \end{equation}
   If \({\left\lVert \mathbf{M}\right\rVert }^2_F = 1\), then the greyness of the FGGCM will possess at least one fixed point.
\end{corollary}

\begin{pf}
Equation \eqref{sig_grey_iter} can equivalently be expressed in matrix form as Eq. \eqref{matrixeq}, with the matrix \(\mathbf{M}\) defined as in Eq. \eqref{sig_M}. This formulation highlights the role of \(\mathbf{M}\) in capturing the relationships governing the greyness iteration process.
\begin{equation}
  \label{matrixeq}
  \bm{A^\circ} =\mathrm{diag}  (\mathbf{M} \cdot \max_{1\leqslant i\leqslant n} (\mathbf{w_{i}^\circ,{A^\circ}})^{^\top})
\end{equation}

If for all \(i\) and \(j\), the condition \(\max (w_{ij}^\circ, A_j^\circ) = A_j^\circ\) is satisfied, then Eq. \eqref{matrixeq} simplifies to Eq. \eqref{simplified_eq}. 
\begin{equation}
  \label{simplified_eq}
  \bm{A^\circ} =\mathbf{M} \cdot  \bm{A}{^\circ}^{\top}
\end{equation}
Thus, the following relationship holds:
\begin{equation*}
  \begin{split}
    &\sum_{i=1}^{n} \sum_{j=1}^{n}\left(\frac{\hat{A}_{i}\left|{\hat{A}_{j} \hat{w}_{ij}}\right|\theta\left(C_j^\circ - \hat{w}_{ij}^\circ\right) }{\sum_{j=1}^{n}\left|\hat{w}_{ij}\hat{A}_j\right|}\right)^2 \\
    = & \sum_{i=1}^{n} \sum_{j=1}^{n}\left(\frac{\hat{A}_{i}\left|{\hat{A}_{j} \hat{w}_{ij}}\right| }{\sum_{j=1}^{n}\left|\hat{w}_{ij}\hat{A}_j\right|}\right)^2 = {\left\lVert \mathbf{M}\right\rVert }^2_F
  \end{split}
\end{equation*}

Hence, when the kernel of the FGGCM converges to a fixed point and the sigmoid function serves as the activation function, if the condition $\max (\bm{w_{i}^\circ, \bm{A}^\circ}) = \bm{A}^\circ$ is fulfilled, it follows that if ${\left\lVert \mathbf{M} \right\rVert}_F < 1$, the greyness of the FGGCM will necessarily converge to a unique fixed point. This fixed point corresponds to an eigenvector associated with the eigenvalue of $1$ of the matrix $\mathbf{M}$. Furthermore, based on Lemma \ref{Browder}, if ${\left\lVert \mathbf{M} \right\rVert}_F = 1$, the greyness of the FGGCM will converge to a fixed point, which is also an eigenvector associated with the eigenvalue of $1$ of $\mathbf{M}$. However, in this case, the uniqueness of the fixed point cannot be assured.
\end{pf}

\section{Experiments} \label{sec:result}

The verification process begins by integrating greyness into the Web Experience FCM, transforming it first into the FGCM and subsequently into the FGGCM. By maintaining consistent levels of greyness across these models, their convergence properties are analyzed and compared. This comparative analysis entails a concurrent evaluation of Theorems \ref{thmfcm}, \ref{thmfgcm}, and \ref{thmFGGCMsig}. The objective is to establish that Theorems \ref{thmfcm} and \ref{thmfgcm} represent specific cases within the more comprehensive framework defined by Theorem \ref{thmFGGCMsig}.

Apply Eqs. \eqref{addgreyl} and \eqref{addgreyu} to incorporate greyness into the structure of Matrix \eqref{wweb}. 
\begin{equation}
  \label{addgreyl}
  \underline{w_{ij}} = \left\{\begin{array} {c c} w_{ij} - g^\circ & \mathrm{if} \,\, w_{ij} - g > -1 \\ 
    -1 & \mathrm{if} \,\, w_{ij} - g^\circ \leqslant -1   \\ \end{array} \right.
\end{equation}

\begin{equation}
  \label{addgreyu}
  \overline{w_{ij}} = \left\{\begin{array} {c c} w_{ij} + g^\circ & \mathrm{if} \,\, w_{ij} + g \leqslant 1 \\ 
    1 & \mathrm{if} \,\, w_{ij} + g^\circ > 1   \\ \end{array} \right.
\end{equation}

In Eqs. \eqref{addgreyl} and \eqref{addgreyu}, the parameter \( g^\circ > 0, g \in \mathbb{R} \) serves as a mechanism to regulate the intensity of the greyness incorporated into the matrix. To guarantee the applicability of Theorem \ref{thmfgcm}, for Matrix \eqref{wweb}, greyness should not be introduced using Eqs. \eqref{addgreyl} and \eqref{addgreyu} when \( \left\lvert w_{ij} \right\rvert < g \). This approach ensures that the conditions \( \underline{w_{ij}} \leqslant \overline{w_{ij}} \leqslant 0 \) and \( 0 \leqslant \underline{w_{ij}} \leqslant \overline{w_{ij}} \) are consistently satisfied. By setting \( g = 0.01 \), the corresponding weight matrix in IGN form is given by Eq. \eqref{wwebgreyav}.

\begin{figure*}[htbp]
  \begin{equation}
    \small
    \setlength{\arraycolsep}{-1pt}
    \begin{split}
      &\otimes \mathbf{W} =\\
    &\begin{pmatrix}  
      [0,0]  &[-0.91,-0.89]  &[-0.89,-0.87]  &[0.99,1.00]  &[-0.86,-0.84]  &[-0.84,-0.82]  &[0.99,1.00]  \\ 
    [0.99,1.00]  &[0,0]  &[-0.94,-0.92]  &[-0.90,-0.88]  &[-0.91,-0.89]  &[-0.95,-0.93]  &[0.99,1.00]  \\ 
    [-0.99,-0.97]  &[-0.94,-0.92]  &[-1.00,-0.99]  &[-1.00,-0.99]  &[0.99,1.00]  &[0.99,1.00]  &[0.99,1.00]  \\ 
    [-1.00,-0.98]  &[-0.90,-0.88]  &[-1.00,-0.99]  &[-0.40,-0.38]  &[0.72,0.74]  &[0.57,0.59]  &[0.69,0.71]  \\ 
    [0.99,1.00]  &[0.99,1.00]  &[0.99,1.00]  &[0.99,1.00]  &[-0.81,-0.79]  &[0.50,0.52]  &[0.99,1.00]  \\ 
    [0.99,1.00]  &[0.99,1.00]  &[0.82,0.84]  &[0.99,1.00]  &[0.50,0.52]  &[-0.40,-0.38]  &[0.99,1.00]  \\ 
    [0.99,1.00]  &[0.99,1.00]  &[0.99,1.00]  &[0.99,1.00]  &[-0.72,-0.70]  &[-0.50,-0.48]  &[-0.68,-0.66]  \\ 
    \end{pmatrix}
    \end{split}
    \label{wwebgreyav}
  \end{equation}
\end{figure*}

Eq. \eqref{wwebgreyav} meets the requirements for determining \( \mathbf{W}^{*} \) as outlined in Theorem \ref{thmfgcm}. Utilizing Eq. \eqref{wstar}, the computation of \( \mathbf{W}^{*} \) yields the result presented in Matrix \eqref{wstarweb}. 
\begin{equation}
  \label{wstarweb}
  \mathbf{W}^{*}= \begin{pmatrix}  
    0  &0.91  &0.89  &1.00  &0.86  &0.84  &1.00  \\ 
  1.00  &0  &0.94  &0.90  &0.91  &0.95  &1.00  \\ 
  0.99  &0.94  &1.00  &1.00  &1.00  &1.00  &1.00  \\ 
  1.00  &0.90  &1.00  &0.40  &0.74  &0.59  &0.71  \\ 
  1.00  &1.00  &1.00  &1.00  &0.81  &0.52  &1.00  \\ 
  1.00  &1.00  &0.84  &1.00  &0.52  &0.40  &1.00  \\ 
  1.00  &1.00  &1.00  &1.00  &0.72  &0.50  &0.68  \\ 
  \end{pmatrix}
\end{equation}

To transform Eq. \eqref{wwebgreyav} into the GGN form as illustrated in Eq. \eqref{wwebggn}, the greyness values are appropriately integrated into the matrix representation.

  \begin{equation}
    \label{wwebggn}
    \setlength{\arraycolsep}{2pt}
     \mathbf{W}_{web}^\pm = 
    \begin{pmatrix}  
      0.000_{0.000} &-0.900_{0.010} &-0.880_{0.010} &0.995_{0.005} &-0.850_{0.010} &-0.830_{0.010} &0.995_{0.005} \\ 
    0.995_{0.005} &0.000_{0.000} &-0.930_{0.010} &-0.890_{0.010} &-0.900_{0.010} &-0.940_{0.010} &0.995_{0.005} \\ 
    -0.980_{0.010} &-0.930_{0.010} &-0.995_{0.005} &-0.995_{0.005} &0.995_{0.005} &0.995_{0.005} &0.995_{0.005} \\ 
    -0.990_{0.010} &-0.890_{0.010} &-0.995_{0.005} &-0.390_{0.010} &0.730_{0.010} &0.580_{0.010} &0.700_{0.010} \\ 
    0.995_{0.005} &0.995_{0.005} &0.995_{0.005} &0.995_{0.005} &-0.800_{0.010} &0.510_{0.010} &0.995_{0.005} \\ 
    0.995_{0.005} &0.995_{0.005} &0.830_{0.010} &0.995_{0.005} &0.510_{0.010} &-0.390_{0.010} &0.995_{0.005} \\ 
    0.995_{0.005} &0.995_{0.005} &0.995_{0.005} &0.995_{0.005} &-0.710_{0.010} &-0.490_{0.010} &-0.670_{0.010} \\ 
    \end{pmatrix}
  \end{equation}
  
The FGCM input vector is represented as:
\begin{equation*}
  \begin{split}
    \left([0.99,1.00],[0.99,1.00],[0.99,1.00],[0.99,1.00],
    [0.99,1.00],[0.99,1.00],[0.00,0.00] \right)
  \end{split}
\end{equation*} 

The corresponding GGN form is expressed as:

\begin{equation}
  \label{webinputggn}
\begin{split}
    \left(0.995_{(0.010)},0.995_{(0.010)},0.995_{(0.010)},0.995_{(0.010)},
    0.995_{(0.010)},0.995_{(0.010)},0.000_{(0.000)}\right).
\end{split}
\end{equation}

\begin{figure}[htbp]
  \centering
    \includegraphics[width=0.6\linewidth]{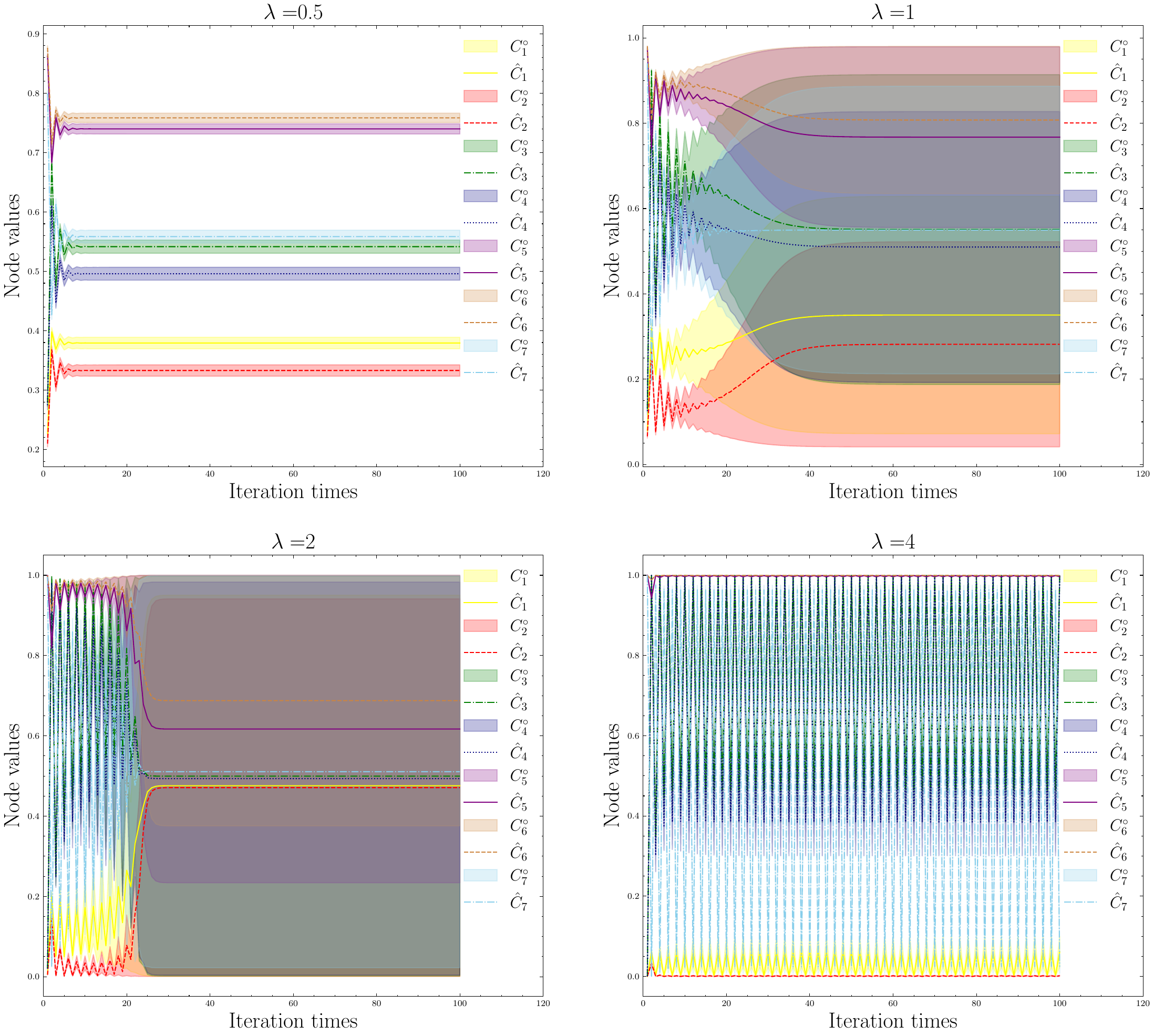}
  \caption{The results generated by the Web Experience FGCM for various values of \(\lambda\).}
  \label{figFGCM_web}
\end{figure}

\begin{figure}[htbp]
  \centering
    \includegraphics[width=.6\linewidth]{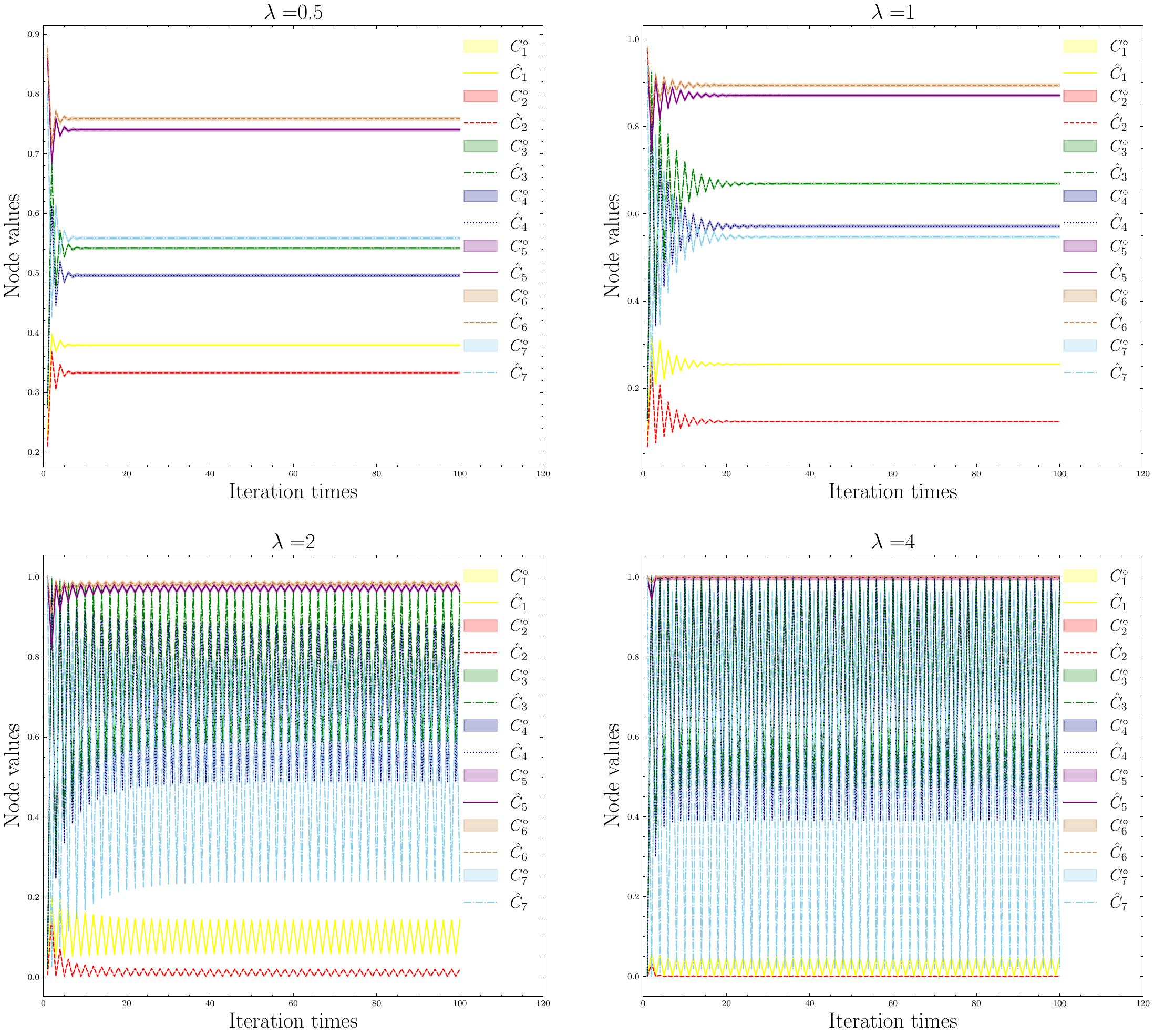}
  \caption{The results produced by the Web Experience FGGCM across different values of \(\lambda\).}
  \label{figFGGCM_web}
\end{figure}

Based on the matrices \eqref{wwebgreyav} and \eqref{wwebggn}, the reasoning results for the FGCM and FGGCM are presented in Figs \ref{figFGCM_web} and \ref{figFGGCM_web}, respectively. These figures illustrate the convergence behavior and the impact of greyness incorporation within the models, highlighting the dynamics under the respective frameworks.

The following elaborates on specific cases where Theorem \ref{thmFGGCMsig} can determine convergence. These scenarios represent instances where the earlier Theorems \ref{thmfcm} and \ref{thmfgcm} are unable to provide a definitive assessment.

\begin{itemize}
  \item Case 1: One significant limitation of Theorem \ref{thmfgcm} lies in its inability to handle cases where the weight matrix contains elements spanning both negative and positive values, such as \([-a, +b]\) with \(a, b > 0\) and \(a, b \in \mathbb{R}\). This limitation arises because the computation of \(\mathbf{W}^*\) becomes infeasible under these conditions, rendering the theorem unsuitable for such scenarios. In contrast, the newly proposed Theorem \ref{thmFGGCMsig} addresses this issue, thereby extending its applicability to weight matrices with mixed elements in the range \([-a, +b]\). In this case, set \(\otimes w_{{11}} = [-0.1, 0.1]\), the calculation of \(\mathbf{W}^*\) is not possible; however, \(\hat{\mathbf{W}}\) can still be obtained, with \(w_{{11}}^\pm\) determined as \(0_{0.1}\). The convergence behavior of FGCM and FGGCM under these conditions is illustrated in Figs \ref{figFGCM_web_cw} and \ref{figFGGCM_web_cw}, respectively.
  
  \begin{figure}[htbp]
    \centering
      \includegraphics[width=.6\linewidth]{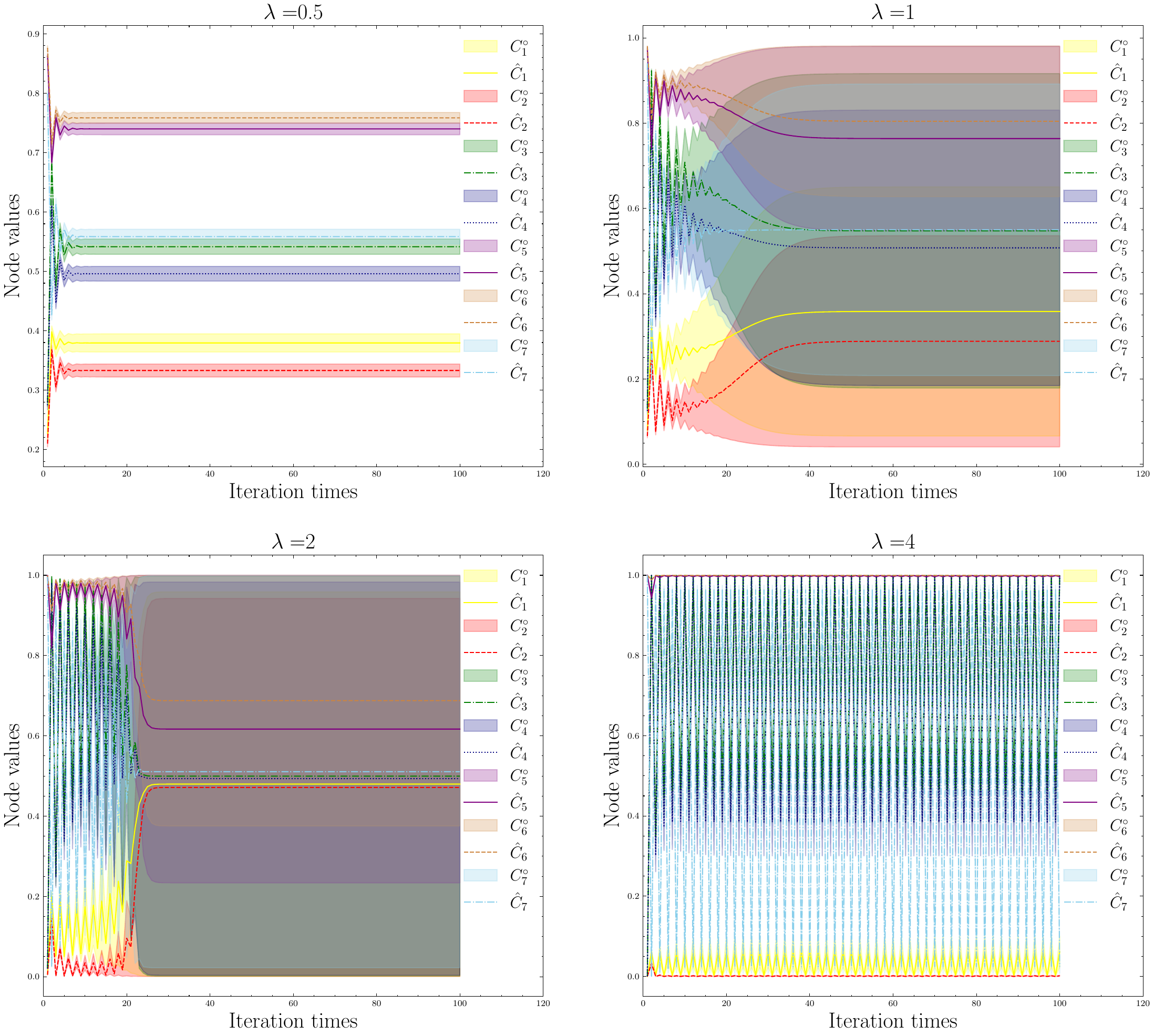}
    \caption{The output of the Web Experience FGCM under various \(\lambda\) values, incorporating \(\otimes w_{{11}} = [-0.1, 0.1]\).}
    \label{figFGCM_web_cw}
  \end{figure}
  \begin{figure}[htbp]
    \centering
      \includegraphics[width=.6\linewidth]{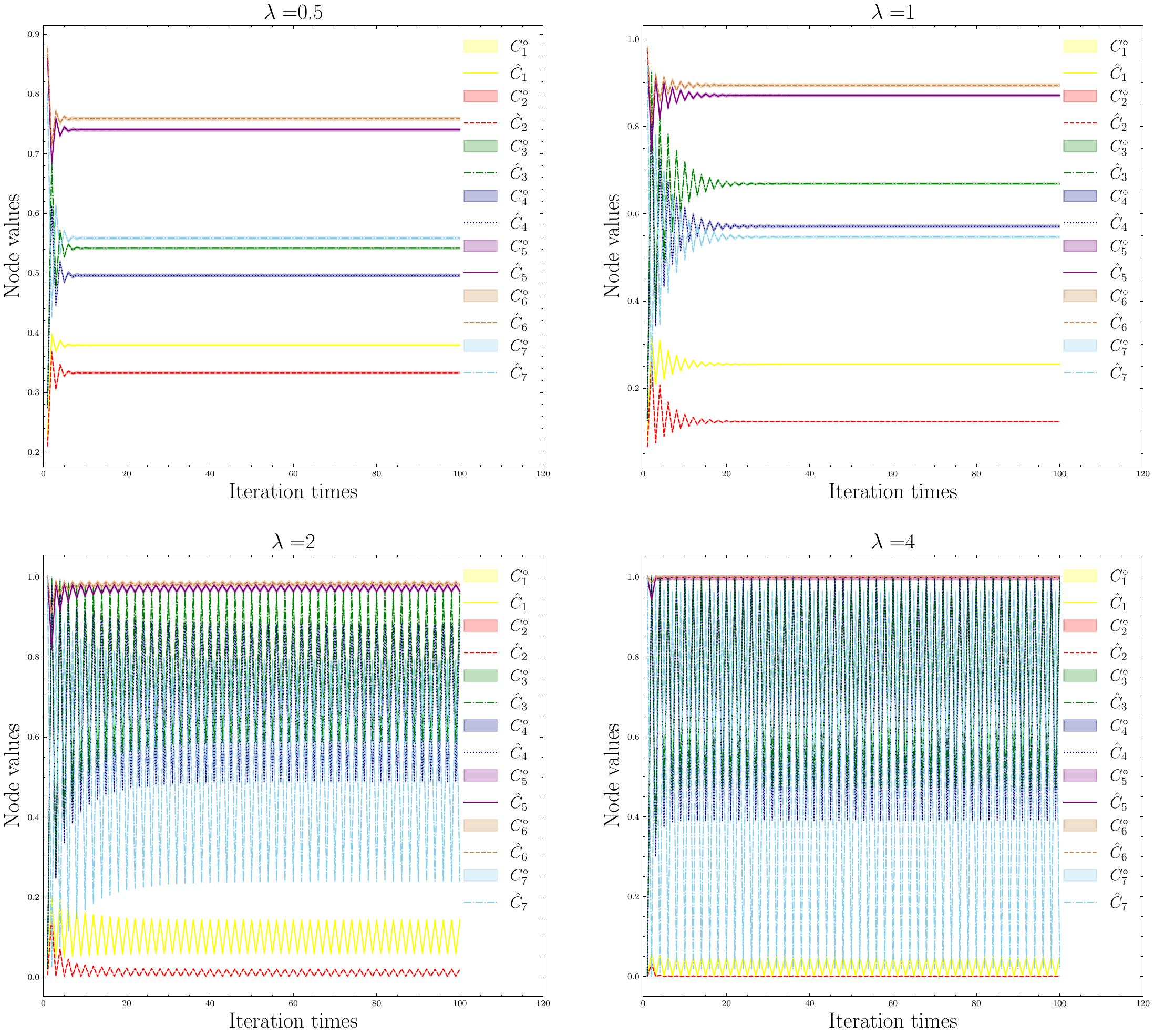}
    \caption{The output of the Web Experience FGGCM for varying \(\lambda\) values, with \(w_{{11}}^\pm = 0_{0.1}\).}
    \label{figFGGCM_web_cw}
  \end{figure}

\item Case 2: Beyond managing weight matrices containing elements of the form \([-a, +b]\), FGGCM must also be evaluated with weights incorporating more complex representations of GGNs. The capability to reason with such intricate weight structures is a defining advantage of FGGCM over FGCM. When \(\mathbf{W}^\pm\) includes multiple IGNs or fuzzy numbers, conventional FCM and FGCM are unable to operate due to strict data format limitations. As a result, Theorem \ref{thmfgcm} cannot be employed to assess system convergence under these conditions. To overcome these limitations, FGGCM provides the necessary framework to handle complex scenarios effectively. Additionally, Theorem \ref{thmFGGCMsig} offers a comprehensive approach for determining system convergence in the presence of such sophisticated weight structures. This framework ensures accurate convergence analysis even when dealing with intricate data formats. Consider the following GGN-based weight definitions:  
\(w_{{11}}^\pm = [-0.9, -0.75] \cup [0.4, 0.9]\),  
\(w_{{12}}^\pm = [-0.95, -0.89] \cup -0.83 \cup [-0.8, -0.75]\),  
\(w_{{33}}^\pm = [-1, -0.95] \cup [-0.94, -0.90] \cup [-0.89, 0.88]\),  
\(w_{{15}}^\pm = [0.99, 1] \cup [0.95, 0.98] \cup [-0.90, 0.93]\).
In such cases, reasoning can only be conducted using FGGCM, as other frameworks are inadequate. The results of applying FGGCM to this configuration are depicted in Fig. \ref{figFGGCM_web_complex}.
  
  \begin{figure}[htbp]
    \centering
      \includegraphics[width=.6\linewidth]{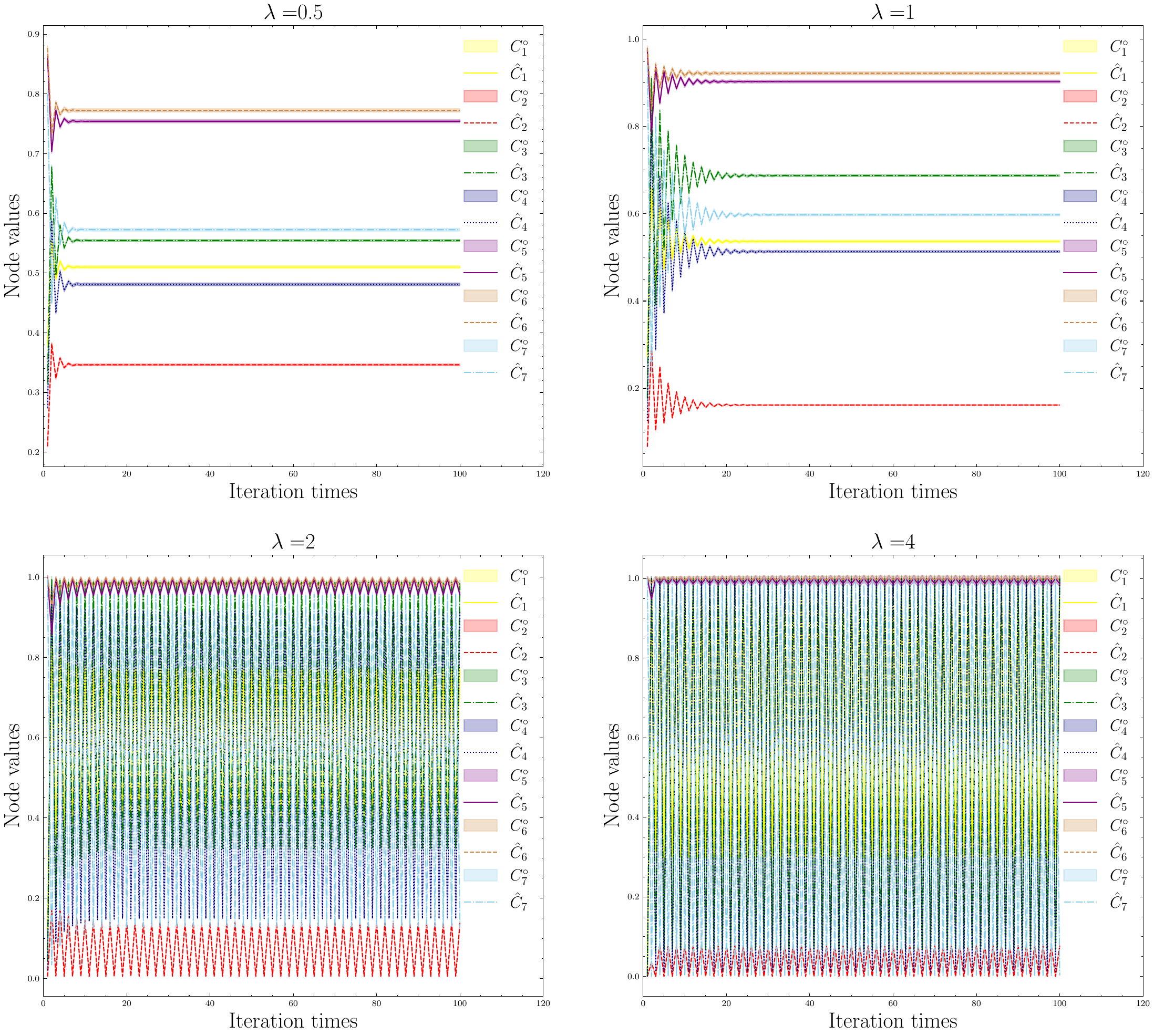}
    \caption{The output of the Web Experience FGGCM for various \(\lambda\) values, utilizing more complex weights.}
    \label{figFGGCM_web_complex}
  \end{figure}
\end{itemize}

\begin{figure}[htbp]
  \centering
    \includegraphics[width=.6\linewidth]{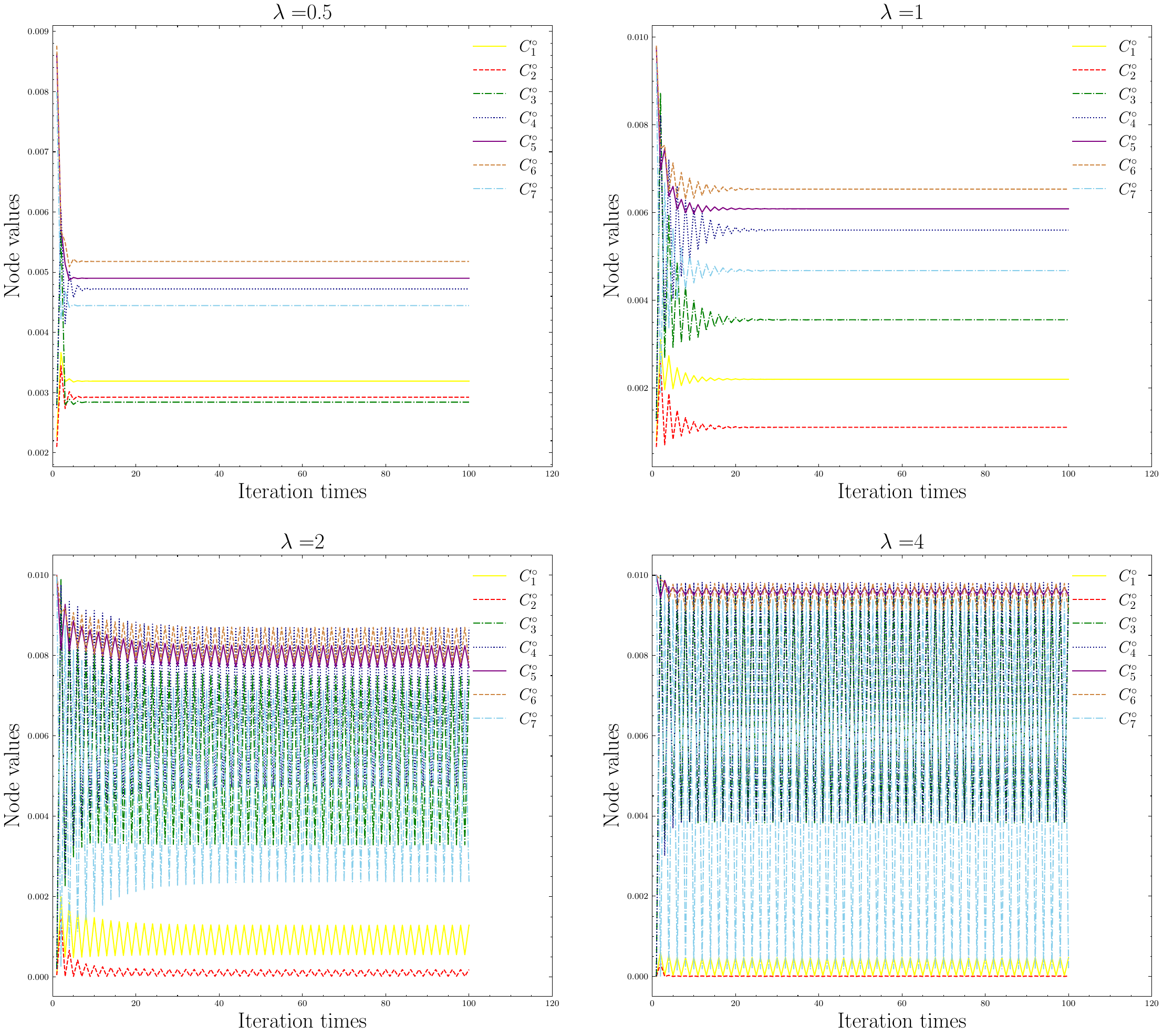}
  \caption{The variation in greyness of the Web Experience FGGCM across different values of \(\lambda\).}
  \label{figFGGCM_web_g}
\end{figure}

Existing literature lacks specific conditions addressing the convergence of greyness in FGGCM. This paper bridges this gap by proposing a condition that explicates the greyness convergence behavior within FGGCM. Using the weight matrix given in Eq. \eqref{wwebggn} and the input vector defined in Eq. \eqref{webinputggn}, the iterative variation of greyness in the Web Experience FGGCM can be analyzed. This progression is depicted in Fig. \ref{figFGGCM_web_g}.

It is observed that for \(\lambda = 0.5\) and \(\lambda = 1\), the greyness in FGGCM converges to a fixed value. In contrast, for \(\lambda = 2\) and \(\lambda = 4\), the greyness does not exhibit convergence but instead oscillates periodically, corresponding to the periodic oscillations of the kernels.

\section{Discussion}   \label{sec:discussion}

This section focuses on analyzing the simulation results, which are utilized to substantiate the theorems proposed in this study. The objective is to offer empirical validation for the theoretical claims, thereby strengthening the practical relevance of these theorems within the framework of FGGCM.

\subsection{The Convergence of Kernels}

The convergence behavior of the Web Experience FCM, FGCM, and FGGCM, utilizing the sigmoid function as the activation function, is analyzed under varying \(\lambda\) values. To assess the convergence of these models, the \(\left\lVert \mathbf{W} \right\rVert _F \times \lambda\) values are computed based on Theorems \ref{thmfcm}, \ref{thmfgcm}, and \ref{thmFGGCMsig}, with the results presented in Table \ref{webcompare}. When \(\left\lVert \mathbf{W} \right\rVert _F \times \lambda < 4\), it indicates that the kernels of the respective FCM, FGCM, and FGGCM models converge to a unique fixed point.

\begin{table}[htbp]
  \caption{In the context of Web engineering, the corresponding values of $\left\lVert \mathbf{W}\right\rVert _F \times \lambda$ for various \(\lambda\) settings.
  }\label{webcompare}
  \centering
  \begin{tabular*}{.6\linewidth}{ccccc}
    \toprule    
     & $\lambda = 0.5$ &$\lambda = 1$& $\lambda = 2$ & $\lambda = 4$ \\ 
    \midrule 
    $\left\lVert \mathbf{W}\right\rVert _F \times \lambda$& \textbf{3.0680}  &6.1359  &12.2719  &24.5437  \\ 
    $\left\lVert \mathbf{W}^*\right\rVert _F \times \lambda$&\textbf{3.0829}  &6.1657  &12.3315  &24.6630  \\ 
    $\left\lVert \hat{\mathbf{W}}\right\rVert _F \times \lambda$&\textbf{3.0586} &6.1172  &12.2344  &24.4688  \\ 
    $\left\lVert \hat{\mathbf{W}}_{1}\right\rVert _F \times \lambda$&\textbf{3.0586} &6.1172  &12.2344  &24.4688  \\ 
    $\left\lVert \hat{\mathbf{W}}_{{mc}}\right\rVert _F \times \lambda$&\textbf{3.0186}   &6.0372  &12.0745  &24.1489  \\ 
    \bottomrule    
    \end{tabular*}
  \end{table}

In Table \ref{webcompare}, \(\mathbf{W}_{1}\) denotes a modified weight matrix derived from Eq. \eqref{wwebgreyav} by replacing the element \(\otimes w_{11}\) with the interval \([-0.1, 0.1]\), while \(\mathbf{W}_{{mc}}\) represents a weight matrix incorporating multiple distinct IGNs or fuzzy numbers. In these specific cases, the convergence can only be determined using the theorems introduced in this study. The bolded values in the table highlight instances where the kernels of FGGCM are guaranteed to converge to a unique fixed point. This assertion is corroborated by the convergence behaviors depicted in Figs \ref{figfcmweb}, \ref{figFGCM_web}, \ref{figFGGCM_web}, \ref{figFGCM_web_cw}, \ref{figFGGCM_web_cw}, and \ref{figFGGCM_web_complex}. 

It is important to observe that in certain scenarios where \(\left\lVert \mathbf{W}\right\rVert _F \times \lambda > 4\), the FCM, FGCM, and FGGCM still exhibit convergence tendencies. This outcome arises from the fact that Theorems \ref{thmFGGCMsig}, \ref{thmfcm}, and \ref{thmfgcm} provide sufficient, but not necessary conditions for determining the convergence of these models. Consequently, practical applications require a comprehensive evaluation of convergence by integrating specific circumstances and the conditions outlined in the theorems.

By examining Figs \ref{figfcmweb}, \ref{figFGCM_web}, and \ref{figFGGCM_web}, the reasoning processes of the Web Experience FCM, FGCM, and FGGCM align with the descriptions provided in \cite{Chen2020}. Notably, FGGCM demonstrates strong compatibility with the reasoning results of FGCM and FCM, while also exhibiting convergence properties akin to those of FCM. This behavior can be attributed to the structural similarity between FGGCM and FCM in their kernel operations. Moreover, FGGCM effectively minimizes the amplification of greyness during kernel computation, which represents a key advantage of this model. In contrast, FGCM experiences an unavoidable amplification of greyness throughout the iterative process. As a result, while FGGCM and FCM may enter a limit cycle, FGCM often progresses toward a fixed point under certain conditions (\(\lambda = 1\) and \(2\)) due to the increasing greyness in its calculations. Importantly, this phenomenon does not contradict the conditions set forth by Theorem \ref{thmfgcm}, as the theorem does not guarantee convergence when \(\left\lVert \mathbf{W}^*\right\rVert _F \times \lambda > 4\).

Among the experiments, particular emphasis is placed on the scenario where the weight matrix of FGGCM includes an interval of the form \([-a, +b]\), specifically with \(\otimes w_{11} = [-0.1, 0.1]\). In this case, the weights do not meet the conditions \(\underline{w_{ij}} \leqslant \overline{w_{ij}} \leqslant 0\) or \(0 \leqslant \underline{w_{ij}} \leqslant \overline{w_{ij}}\) outlined in Theorem \ref{thmfgcm}, thereby rendering it inapplicable for directly assessing the convergence of FGCM. However, by transforming the IGN \([-a, +b]\) into the simplified GGN, FGGCM can be employed for reasoning, and Theorem \ref{thmFGGCMsig} can be utilized to evaluate its convergence. The reasoning results, presented in Fig. \ref{figFGGCM_web_cw}, align with the convergence criteria specified by Theorem \ref{thmFGGCMsig}.

A closer comparison between Fig. \ref{figFGCM_web_cw} and Fig. \ref{figFGGCM_web_cw} reveals reasoning results that are consistent with those observed in Fig. \ref{figFGCM_web} and Fig. \ref{figFGGCM_web}. For \(\lambda = 1\) and \(2\), FGCM exhibits an inevitable amplification of greyness during its iterative process. Specifically, the increased greyness of \(\otimes w_{11}\) directly contributes to the rising greyness of \(C_1\), as evidenced by the comparison between Fig. \ref{figFGCM_web} and Fig. \ref{figFGCM_web_cw}.

Finally, as shown in Fig. \ref{figFGGCM_web_complex}, when the weights exhibit a more intricate GGN structure, their convergence can only be assessed using Theorem \ref{thmFGGCMsig}. A review of Table \ref{webcompare} alongside Fig. \ref{figFGGCM_web_complex} indicates that the experimental results align with and validate the conclusions drawn from Theorem \ref{thmFGGCMsig}.

\subsection{The Convergence of Greyness}

Fig. \ref{figFGGCM_web_g} illustrates the variations in greyness for the Web Experience FGGCM employing a sigmoid activation function. Analysis of the figure reveals that for \(\lambda\) values of \(0.5\) and \(1\), the greyness of FGGCM demonstrates a convergent behavior, indicating that the model stabilizes towards a fixed greyness value under these parameter settings. Conversely, when \(\lambda\) increases to \(2\) or \(4\), a notable shift occurs. In these cases, the kernel of FGGCM ceases to converge and instead exhibits periodic oscillations. This results in the greyness failing to settle at a fixed point and instead fluctuating periodically within a defined range, driven by the oscillatory nature of the kernel.

The convergence of greyness in the Web Engineering FGGCM can be evaluated using Eq. \eqref{siggreynessconvergence} from Theorem \ref{thmFGGCMsig}. Define  
$$  
\frac{\hat{A}_{i}\left|{\hat{A}_{j} \hat{w}_{ij}}\right|\theta\left(C_j^\circ - \hat{w}_{ij}^\circ\right) }{\sum_{j=1}^{n}\left|\hat{w}_{ij}\hat{A}_j\right|} = \widetilde{m}_{ij},  
$$  
where \(\widetilde{m}_{ij}\) represents an element of the matrix \(\mathbf{\widetilde{M}}\). The calculated $\left\lVert \widetilde{\mathbf{M} }\right\rVert _F  $ values under various \(\lambda\) settings are summarized in Table \ref{mfweb}.

\begin{table}[htbp]
  \caption{The $\left\lVert \widetilde{\mathbf{M} }\right\rVert _F  $ values for the Web Experience FGGCM across various \(\lambda\) settings.
  }\label{mfweb}
  \centering
  \begin{tabular*}{.5\linewidth}{ccccc}
    \toprule    
    & $\lambda = 0.5$ &$\lambda = 1$& $\lambda = 2$ & $\lambda = 4$ \\ 
    \midrule 
    $\left\lVert \widetilde{\mathbf{M} }\right\rVert _F  $ & 0.1984 &0.3466  &0.5217  &0.6076  \\ 
    \bottomrule    
    \end{tabular*}
  \end{table}

When \(\lambda = 0.5\) or \(1\), the condition \(\left\lVert \widetilde{\mathbf{M}} \right\rVert _F < 1\) is satisfied, and under these circumstances, the kernel of FGGCM converges to a fixed point. Consequently, it can be concluded that the greyness of FGGCM also converges to a unique fixed point, leading to the overall convergence of FGGCM to a unique fixed point as per Theorem \ref{thmFGGCMsig}. However, for \(\lambda = 2\) or \(4\), despite satisfying \(\left\lVert \widetilde{\mathbf{M}} \right\rVert _F < 1\), the kernels of FGGCM fail to converge, thus violating the conditions specified in Corollary \ref{sig_grey_convergence}. In this case, the proposed theorems cannot determine the greyness convergence. Analyzing Eq. \eqref{sig_grey_iter}, it becomes evident that when the sigmoid function serves as the activation function, the greyness of FGGCM is affected by the kernels of two consecutive iteration steps. This dependency implies that the greyness is influenced not only by the current state of the kernel but also by its previous state.  
If the kernels of FGGCM form a limit cycle, the greyness is unlikely to converge to a fixed point. The periodic nature of the kernel states results in corresponding periodic variations in greyness during the iteration process, preventing it from stabilizing. This phenomenon highlights the dynamic complexity of FGGCM under certain conditions, particularly when the sigmoid function is employed as the activation function.

When the greyness of FGGCM is set to \(0\), it becomes evident that Eq. \eqref{sigkernelconvergence} reduce to and Eq. \eqref{eqfcmsig}.This observation indicates that Theorems \ref{thmfcm} and \ref{thmfgcm} can be regarded as specific instances of Theorems \ref{thmFGGCMsig}. Such a result aligns with the interpretation of FCM and FGCM as particular cases of FGGCM, thereby establishing a theoretical unification of these models and shedding light on their underlying interrelationships.

\section{Conclusions}   \label{sec:conclusion}

This paper investigates the convergence properties of FGGCMs with the sigmoid activation function. By employing theorems such as the Banach Fixed Point Theorem and the Browder-Gohde-Kirk Fixed Point Theorem, the study establishes sufficient conditions for the kernels and greyness of FGGCM to converge to fixed points. Through rigorous theoretical derivations and comprehensive simulations, the results confirm that the proposed conditions effectively predict the convergence behavior of FGGCMs under various parameter settings. The findings further reveal the intrinsic connections between FCM, FGCM, and FGGCM, with the former two models shown as special cases of the latter. Future research will focus on extending these results to other activation functions, exploring convergence in complex spaces, analyzing the occurrence of limit cycles or chaotic states, and improving computational efficiency and learning algorithms for FGGCMs in practical applications.

\section*{Declaration of Competing Interest}

The authors declare that they have no known competing financial interests or personal relationships that could have appeared to influence the work reported in this paper.

\section*{Acknowledgments}

This work was supported in part by the 
National Natural Science Foundation of China under Grant 61305133, 61876187, 52372398.
\bibliography{convergence_s}
\end{document}